\documentclass[aps,prb,twocolumn,amsmath,amssymb,floatfix,showpacs]{revtex4-1}

\usepackage{graphicx}
\usepackage{dcolumn}
\usepackage{bm}
\usepackage{multirow}
\usepackage{color}
\usepackage{epstopdf}

\newcommand{\bk}{\mathbf{k}}
\newcommand{\bp}{\mathbf{p}}
\newcommand{\dist}{g}
\newcommand{\dn}{\downarrow}
\newcommand{\Ed}{\veps_d}
\newcommand{\Edz}{\veps_{d0}}
\newcommand{\half}{{\textstyle\frac{1}{2}}}
\newcommand{\imp}{\mathrm{imp}}
\newcommand{\nz}{\hat{\mathbf{n}}_0}
\newcommand{\loc}{\mathrm{loc}}
\newcommand{\pdag}{\phantom{\dag}}
\newcommand{\reg}{\mathrm{reg}}
\newcommand{\tbeta}{\tilde{\beta}}
\newcommand{\tdelta}{\tilde{\delta}}
\newcommand{\tphi}{\tilde{\phi}}
\newcommand{\tgamma}{\tilde{\gamma}}
\newcommand{\tx}{\tilde{x}}
\newcommand{\up}{\uparrow}
\newcommand{\veps}{\varepsilon}

\begin{document}
\title{Critical charge fluctuations in a pseudogap Anderson model}

\author{Tathagata Chowdhury}
\email{Electronic address: tatha@phys.ufl.edu}
\author{Kevin Ingersent}
\affiliation{Department of Physics, University of Florida, P.O.\ Box 118440,
Gainesville, Florida 32611--8440, USA}

\date{\today}

\begin{abstract}
The Anderson impurity model with a density of states $\rho(\veps) \propto
|\veps|^r$ containing a power-law pseudogap centered on the Fermi energy
($\veps = 0$) features for $0<r<1$ a Kondo-destruction quantum critical point
(QCP) separating Kondo-screened and local-moment phases. The observation of
mixed valency in quantum critical $\beta$-YbAlB$_4$ has prompted study of
this model away from particle-hole symmetry. The critical spin response
associated with all Kondo destruction QCPs has been shown to be accompanied,
for $r=0.6$ and noninteger occupation of the impurity site, by a divergence of
the local charge susceptibility on both sides of the QCP. In this work, we
use the numerical renormalization-group method to characterize the
Kondo-destruction charge response using five critical exponents, which are
found to assume nontrivial values only for $0.55\lesssim r < 1$. For $0 < r
\lesssim 0.55$, by contrast, the local charge susceptibility shows no
divergence at the QCP, but rather exhibits nonanalytic corrections to a regular
leading behavior. Both the charge critical exponents and the previously obtained
spin critical exponents satisfy a set of scaling relations derived from an
ansatz for the free energy near the QCP. These critical exponents can all be
expressed in terms of just two underlying exponents: the correlation-length
exponent $\nu(r)$ and the gap exponent $\Delta(r)$. The ansatz predicts a
divergent local charge susceptibility for $\nu<2$, which coincides closely with
the observed range $0.55\lesssim r<1$. Many of these results are argued to
generalize to interacting QCPs that have been found in other quantum impurity
models.
\end{abstract}


\maketitle

\section{Introduction}
\label{sec:Intro}

Continuous quantum phase transitions (QPTs) in itinerant electron systems are
conventionally described within a Ginzburg-Landau-Wilson picture of
critical fluctuations of an order parameter characterizing a spontaneously
broken symmetry \cite{Hertz:76,Millis:93,Sachdev:99}. However, experiments on
heavy-fermion metals \cite{vonLohneysen:07+Si:10} have established the
existence of a class of antiferromagnetic quantum critical points (QCPs)
that can be understood only by postulating additional critical modes beyond
order-parameter fluctuations \cite{Coleman:01}. It has been proposed
\cite{Si:01+Si:03} that the additional modes arise from the critical
destruction of the Kondo effect, associated with a jump in the Fermi-surface
volume \cite{Paschen:04,Shishido:05,Friedmann:10} from large in the
paramagnetic phase (where unpaired $f$ electrons are absorbed into Kondo
resonances) to small in the antiferromagnetic phase (where the Kondo
resonances are destroyed and the $f$ electrons are localized).

The picture of critical Kondo destruction was originally developed in
the Kondo limit of integer $f$ occupancy. More recently, the discovery
of unconventional quantum criticality \cite{Nakatsuji:08,Matsumoto:11} in
mixed-valent \cite{Okawa:10} $\beta$-YbAlB$_4$ has prompted interest in
critical Kondo destruction at mixed valence. A toy model for this phenomenon
is the particle-hole-asymmetric Anderson impurity model with a density of
states $\rho(\epsilon) \propto |\epsilon|^r$ that vanishes in power-law
fashion on approach to the Fermi energy $\epsilon = 0$. The model features a
Kondo-destruction QCP separating a strong-coupling (Kondo-screened) phase
from a local-moment (Kondo-destroyed) phase \cite{Gonzalez-Buxton:96,%
Bulla:97,Gonzalez-Buxton:98,Fritz:04}. A study conducted using a combination
of continuous-time quantum Monte Carlo and the numerical renormalization
group (NRG) showed for the particular case $r=0.6$ that Kondo destruction was
accompanied by divergence of a local charge susceptibility on approach to the
QCP from either phase \cite{Pixley:12}. In this case, both spin and charge
responses demonstrate the frequency-over-temperature and magnetic
field-over-temperature scaling characteristic of an interacting QCP.

This paper extends the numerical results provided in Ref.\
\onlinecite{Pixley:12} by determining a complete set of static charge
critical exponents for different values of the band exponent $r$ in the range
$3/8 \lesssim r < 1$ over which the asymmetric pseudogap Anderson model has
an interacting QCP that is distinct from that of its symmetric
counterpart \cite{Gonzalez-Buxton:98,Fritz:04}. We provide a unified
description of both the spin and charge critical behaviors in terms of an
ansatz for the form of the free energy near the QCP, expressing all critical
exponents in terms of just two underlying exponents \cite{Ingersent:02,%
Vojta:06,nomenclature}, which can be termed (in the nomenclature of classical
phase transitions) the ``correlation-length'' exponent $\nu(r)$ and the ``gap''
exponent $\Delta(r)$. The ansatz leads to scaling equations that are obeyed to
high accuracy by numerically determined values of the charge exponents. In
particular, the numerics support a scaling prediction that local charge response
is divergent for $\nu<2$, but regular with nonanalytic corrections for $\nu>2$. 

The outline of the rest of the paper is as follows: Section \ref{sec:background}
defines the pseudogap Anderson Hamiltonian and reviews essential background for
the present work. Our numerical results are presented and interpreted in Sec.\
\ref{sec:results}. Implications of these results for a broader class of quantum
impurity models are discussed in Sec.\ \ref{sec:discuss}.

\section{Background}
\label{sec:background}

\subsection{Model Hamiltonian}
\label{subsec:model}

This work addresses an Anderson model described by the Hamiltonian 
\begin{align}
\label{H}
H
&= \sum_{\bk,\sigma} \veps_{\bk} \, c_{\bk,\sigma}^{\dag}
   c_{\bk,\sigma}^{\pdag}
   + \Ed \, \hat{n}_d + U \hat{n}_{d\up} \, \hat{n}_{d\dn} \notag \\
& \quad  + \: \frac{V}{\sqrt{N_c}} \bigl( d_{\sigma}^{\dag} \,
   c_{\bk,\sigma}^{\pdag} + \mathrm{H.c.} \bigr) + h \hat{S}_{d,z},
\end{align}
where $c_{\bk,\sigma}$ ($d_{\sigma}$) destroys a conduction-band (impurity)
electron with energy $\veps_{\bk}$ ($\Ed$) and spin $z$ component
$\sigma=\pm\half$ (or $\up,\,\dn$),
$\hat{n}_{d\sigma} = d_{\sigma}^{\dag}d_{\sigma}^{\pdag}$ and
$\hat{n}_d=\hat{n}_{d\up}+\hat{n}_{d\dn}$ are number operators, 
$U$ is the Coulomb
interaction between two electrons within the impurity level (taken to be
positive, i.e., repulsive, in our calculations, but see the discussion in
Sec.\ \ref{sec:discuss}), $V$ is the hybridization matrix element between the
impurity level and the on-site linear combination of conduction electrons
(and is assumed without loss of generality to be real and non-negative),
$N_c$ is the number of unit cells in the metallic host and hence the number
of distinct values of $\bk$, and $h$ is a local magnetic field that couples
only to $\hat{S}_{d,z} = \half(\hat{n}_{d\up}-\hat{n}_{d\dn})$, the $z$
component of the impurity spin \cite{units}.

The pseudogap variant of the Anderson model has a density of states
(per unit cell, per spin $z$ orientation)
\begin{equation}
\label{rho}
\rho(\veps)
  = N_c^{-1} \sum_{\bk} \delta(\veps-\veps_{\bk})
  = \rho_0 \, |\veps/D|^r \, \Theta(D-|\veps|),
\end{equation}
where $D$ is the band half width, and $\Theta(x)$ is the Heaviside function.
Values $r>0$ describe a pseudogapped host, while $r=0$ corresponds to a
conventional metal. If $\rho(\veps)$ has unit normalization, then
$\rho_0 = (1+r)/(2D)$. The values of $\rho_0$ and $V$ affect the impurity
properties only in a single combination, the hybridization width
$\Gamma=\pi \rho_0 V^2 \ge 0$.

\subsection{Phase diagram}
\label{subsec:phase}

\begin{figure}[t]
\includegraphics[width=\columnwidth]{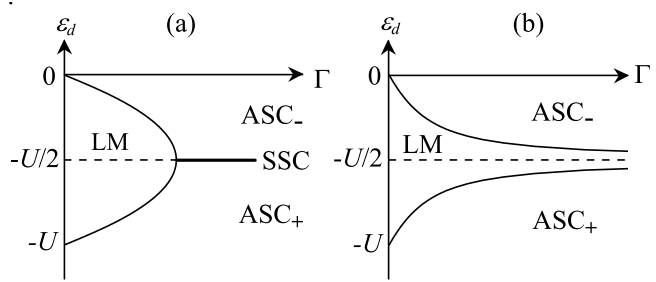}
\caption{\label{fig:phase-diagram}
Schematic phase diagram of the pseudogap Anderson model on the
$\Gamma$-$\varepsilon_d$ plane for fixed $U$ and for band exponents
(a) $0<r<\half$, and (b) $r\ge\half$. In (a), the SSC phase spans just
the solid horizontal line at $\Ed=-U/2$.}
\end{figure}

The phase diagram of the pseudogap Anderson model has been well established
by previous work \cite{Bulla:97,Gonzalez-Buxton:98}. A cut of the phase diagram
on the $\Gamma$-$\Ed$ plane for a fixed value of $U>0$ is shown schematically
for $0<r<\half$ in Fig.\ \ref{fig:phase-diagram}(a) and for
$r\ge\half$ in Fig.\ \ref{fig:phase-diagram}(b).

In the metallic case $r=0$, for all finite values of $U$ and $\Ed$ and for
any $\Gamma>0$, the impurity degree of freedom is completely quenched in the
limit of absolute temperature $T\to 0$, and a single strong-coupling (SC) phase
occupies the entire half space $(U,\,\Ed,\,\Gamma)$ apart from its boundary plane
$\Gamma=0$. Throughout this phase, the impurity contributions to the static spin
susceptibility and the entropy satisfy $\lim_{T\to 0} T\chi_{\imp}=0$ and
$S_{\imp}(T=0)=0$, respectively \cite{units}. The ground-state ``charge'' $Q$,
defined to be the expectation value of the total electron occupancy of the band
and the impurity level measured with respect to half filling, evolves smoothly
from $1$ to $-1$ as $\Ed$ is raised from $-\infty$ to $\infty$.

For $r>0$, by contrast, there is local-moment (LM) phase spanning $-U<\Ed<0$, 
$\Gamma<\Gamma_c(U,\Ed)\equiv\Gamma_c(U,-U-\Ed)$ within which the ground state
contains an unquenched spin degree of freedom characterized by
$\lim_{T\to 0} T\chi_{\imp}=1/4$ and $S_{\imp}(T=0)=\ln 2$.
For $\Gamma>\Gamma_c(U,\Ed)$, the system lies in one of three SC phases.
The symmetric strong-coupling (SSC) phase, reached only for $\Ed=-U/2$
[see solid horizontal line in Fig.\ \ref{fig:phase-diagram}(a)], has
$\lim_{T\to 0} T\chi_{\imp}=r/8$ and $S_{\imp}(T=0)=2r\ln 2$, suggestive of
partial quenching of the impurity spin. The asymmetric strong-coupling phases
ASC$_-$ and ASC$_+$, reached for $\Ed>-U/2$ and $\Ed<-U/2$, respectively, share
the properties $\lim_{T\to 0} T\chi_{\imp}=0$ and $S_{\imp}(T=0)=0$, indicating
complete quenching of the impurity degree of freedom. For $r>0$, the
ground-state charge takes only integer values (in contrast to the case $r=0$):
$Q = 0$ in the LM and SSC phases, $Q=\pm 1$ in the ASC$_{\pm}$ phase.

It should be noted that the SSC phase can be reached only for $0<r<\half$, on
which range $\Gamma_c(U,-U/2)$ is finite [see Fig.\ \ref{fig:phase-diagram}(a)].
For $r\ge\half$, the SSC ground state is unstable, and $\Gamma_c(U,\Ed)$
diverges as $\Ed\to -U/2$ from above or below, so for $\Ed=-U/2$ the system
always lies in the LM phase [see Fig.\ \ref{fig:phase-diagram}(b)].

\subsection{Critical spin response}
\label{subsec:spin-response}

On the boundary between the LM and SC phases, the thermodynamic properties
take values distinct from those in either phase. For example,
$\lim_{T\to 0} T\chi_{\imp}(T) = X(r)$, where $r/8 < X(r) < 1/4$ (see
Fig.\ 14 of Ref.\ \onlinecite{Gonzalez-Buxton:98}). However, the nontrivial
critical properties of the pseudogap Anderson model (and of the pseudogap
Kondo model to which the Anderson model reduces when charge fluctuations on
the impurity site can be neglected) are revealed more clearly in the response
to a local magnetic field that acts solely at the impurity
spin \cite{Ingersent:02,Fritz:04}, as represented by $h$ entering Eq.\ \eqref{H}.
This response is measured by the zero-temperature local magnetization
\begin{equation}
M_{\loc}=-\partial F_{\imp}/\partial h|_{T=0}
=-\langle \hat{S}_{d,z}\rangle|_{T=0}
\end{equation}
and the zero-field local spin susceptibility
\begin{equation}
\chi_s=-\partial^2 F_{\imp}/\partial h^2|_{h=0}
=-\partial\langle \hat{S}_{d,z}\rangle/\partial h|_{h=0},
\end{equation}
where $F_{\imp}$ is the impurity contribution to the system's free energy. 
The value of $M_{\loc}$ in an infinitesimal symmetry-breaking field $h=0^+$ is
nonzero in the LM phase ($\Gamma<\Gamma_c$) but zero in the SC phases
($\Gamma > \Gamma_c$), and therefore serves as an order parameter for the
LM-SC QPT, while the zero-temperature limit of $\chi_s$ diverges on approach
to the QPT from the SC side and is infinite throughout the LM phase.

For $r>1$, $M_{\loc}$ is discontinuous across the phase boundaries, meaning
that the QPTs are first order. For $0<r<1$, by contrast, the QPTs are
continuous and the local magnetic critical behavior can be characterized by a
set of critical exponents $\beta$, $\gamma$, $\delta$, and $x$ defined through
the relations \cite{Ingersent:02,nomenclature}
\begin{subequations}
\label{spin-exponents}
\begin{align}
\label{beta:def}
M_{\loc} (\dist\le 0, h=0^+) &\propto (-\dist)^{\beta},\\
\label{delta:def}
|M_{\loc}(\dist=0)| &\propto |h|^{1/\delta},\\
\label{gamma:def}
\chi_s(T=0, \dist>0) &\propto \dist^{-\gamma},\\
\label{x:def}
\chi_s(\dist=0) &\propto T^{-x},
\end{align}
\end{subequations}
where one can define the nonmagnetic distance to criticality to be
$\dist = \Gamma -\Gamma_0$ at fixed $U=U_0$ and $\Ed=\Edz$ or, alternatively,
$\dist = U_0 - U$ at fixed $\Gamma=\Gamma_0$ and $\Ed=\Edz$, where in either case
$\Gamma_0=\Gamma_c (U_0,\Edz)$.
One can also define a correlation-length exponent $\nu$ via the relation
\begin{equation}
\label{nu:def}
T^* \propto |\dist|^\nu,
\end{equation}
where $T^*$ is a temperature characterizing the crossover from the
quantum-critical regime ($\chi_s\propto T^{-x}$ for $T\gg T^*$) to
either the LM phase ($\chi_s\propto T^{-1}$  for $\dist<0$, $T\ll T^*$)
or one of the SC phases ($\chi_s\simeq\text{const}$ for $\dist>0$,
$T\ll T^*$). Each of the critical exponents $\beta$, $\gamma$, $\delta$,
$x$, and $\nu$ has a nontrivial dependence \cite{Ingersent:02} on the band
exponent $r$.

For $0<r\le r^*\simeq 3/8$, it is found \cite{Ingersent:02} that the critical
exponents take identical values all the way along the phase boundary between
the LM and SC phases. Specifically, particle-hole asymmetry is irrelevant along
the boundary and the QPT is governed by a symmetric QCP. For $r\ge\half$, there
is no QPT at particle-hole symmetry; a single asymmetric QCP governs the
LM-ASC$_-$ boundary, while a second QCP (related to the first by a particle-hole
transformation, and sharing the same set of critical exponents) governs the
LM-ASC$_+$ boundary. Over the range $r^*<r<\half$, there is coexistence of
symmetric and asymmetric QCPs, which have different critical exponents for a
given band exponent $r$.

Over the entire range $0<r<1$, the critical exponents (for both symmetric and
asymmetric QCPs) obey a set of scaling relations \cite{Ingersent:02},
\begin{equation}
\label{spin-scaling}
\beta=\nu(1-x)/2, \quad \gamma=\nu x, \quad 1/\delta=(1-x)/(1+x),
\end{equation}
that are consistent with a scaling ansatz for the critical part of the free
energy,
\begin{equation}
\label{scaling-ansatz}
F_{\imp}^{\text{crit}} = T f \biggl( \frac{\dist}{T^{1/\nu}},
  \frac{|h|}{T^{\Delta/\nu}} \biggr) ,
\end{equation}
written in terms of just two underlying critical exponents, $\nu$
defined in Eq.\ \eqref{nu:def} and the gap exponent $\Delta$.
This scaling form, which is expected to hold only for an interacting QCP
below its upper critical dimension, implies that
\begin{subequations}
\label{ansatz-exponents}
\begin{align}
\label{beta-scaling}
\beta &= \nu - \Delta \, , \\
\label{gamma-scaling}
\gamma &= 2\Delta - \nu \, , \\
\label{delta-scaling}
1/\delta &= \nu/\Delta - 1 \,, \\
\label{x-scaling}
x &= 2\Delta/\nu - 1 \, .
\end{align}
\end{subequations}
Elimination of $\Delta$ from Eqs.\ \eqref{ansatz-exponents} yields
Eqs.\ \eqref{spin-scaling}.

\subsection{Critical charge response}
\label{subsec:charge-response}

Reference \onlinecite{Pixley:12} investigated local charge fluctuations in
the vicinity of the Kondo-destruction QPTs in the pseudogap Anderson model.
The local charge response is the variation of the impurity charge
$\hat{n}_d$, which enters the Hamiltonian with coupling $\Ed$, so it
is natural to define a local charge susceptibility
\begin{equation}
\chi_c = - \partial\langle\hat{n}_d\rangle/\partial\Ed|_{\Ed=\Edz} ,
\end{equation}
near a point $(U_0,\,\Edz,\,\Gamma_0)$ on the phase boundary.
It was reported in Ref.\ \onlinecite{Pixley:12} that $\chi_c$ remains finite
on passage through the particle-hole-symmetric QCPs that occur for $0<r<\half$.
However, it was shown for the specific case $r=0.6$ that $\chi_c$ diverges on
approach to the LM-ASC$_{\pm}$ boundary from either phase. The behavior
for this particular band exponent was found to be described by a pair of
critical exponents $\tgamma$, and $\tx$ defined via the relations
\begin{subequations}
\label{charge-exponents1}
\begin{align}
\label{tgamma:def}
\chi_c(T=h=0) &\propto |\dist|^{-\tgamma},\\
\label{tx:def}
\chi_c(\dist=h=0) &\propto T^{-\tx},
\end{align}
\end{subequations}
where $\dist = U_0 - U$ at fixed $\Ed = \Edz$ and
$\Gamma = \Gamma_0 = \Gamma_c(U_0,\Edz)$.
Equation \eqref{tgamma:def} differs from Eq.\ \eqref{gamma:def} in that
$\chi_c(T=0)$ remains finite for all $\dist<0$ as well as for all $\dist>0$.

\section{Results and Interpretation}
\label{sec:results}

We have systematically extended the results of Ref.\ \onlinecite{Pixley:12}
through study of the particle-hole-asymmetric pseudogap Anderson model with
different values of the band exponent $r$ within the range $r^*<r<1$. We have
departed from Ref.\ \onlinecite{Pixley:12} in that for the most part we have
fixed $U$ and varied $\Ed$ and $\Gamma$, so that we have extracted $\tgamma$
and $\tx$ defined through Eqs.\ \eqref{charge-exponents1} using
$\dist = \Gamma-\Gamma_0$ at fixed $U=U_0$ and $\Ed=\Edz$. For any given $r$,
variation of $U$ and variation of $\Gamma$ are found to yield the same
numerical values of these critical exponents and of other exponents defined
below.

In addition to calculating $\chi_c$, we have also investigated the variation of
the impurity occupancy near the QCP. Since $\langle\hat{n}_d\rangle$ is not
pinned to any fixed value throughout either the LM phase or the ASC$_{\pm}$
phases, it does not act like an order parameter. It proves convenient to define
a zero-temperature local charge
\begin{equation}
Q_{\loc} = \langle\hat{n}_d(U,\,\Ed,\,\Gamma)
   - \hat{n}_d(U_0,\,\Edz,\,\Gamma_0)\rangle|_{T=0} ,
\end{equation}
constructed to vanish at the point $(U_0,\,\Edz,\,\Gamma_0)$ where the phase
boundary is crossed.

We have calculated $Q_{\loc}$ and $\chi_c=\lim_{\Ed\to\Edz}$
$Q_{\loc}/(\Edz-\Ed)$ using the numerical renormalization-group (NRG) method,
as adapted to treat pseudogapped densities of states \cite{Gonzalez-Buxton:98}.
We have employed a discretization parameter $\Lambda=9$, shown in previous NRG
studies of the pseudogap Kondo\cite{Ingersent:02} and Anderson\cite {Pixley:12}
models to yield critical exponents very close to their values in the continuum
limit $\Lambda\to 1$, and retained up to 600 many-body eigenstates after each
NRG iteration. All results \cite{units} shown below are for a representative
point on the LM-ASC$_-$ phase boundary at $U_0=0.1D$, $\Edz=-0.03D$, and
$\Gamma_0=\Gamma_c(U_0,\,\Edz)$. However, other runs indicate that exponents
depend on $r$ but not on the specific values of $U_0$, $\Edz$, and $\Gamma_0$
(provided that $\Edz\ne-U_0/2$).

\begin{figure}
\includegraphics[width=\columnwidth]{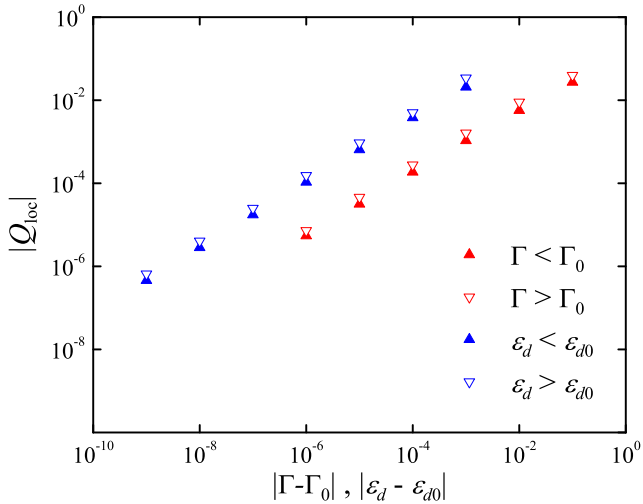}
\caption{\label{fig:power-laws} (Color online)
$|Q_{\loc}|$ vs distance from the phase boundary along the $\Gamma$ and $\Ed$
axes for $r=0.6$. Filled (hollow) symbols represent points in the LM (ASC$_-$)
phase. The linear variations on this log-log plot indicate power-law behavior
in accordance with Eqs.\ \eqref{tbeta:def}.}
\end{figure}

\begin{figure}
\includegraphics[width=\columnwidth]{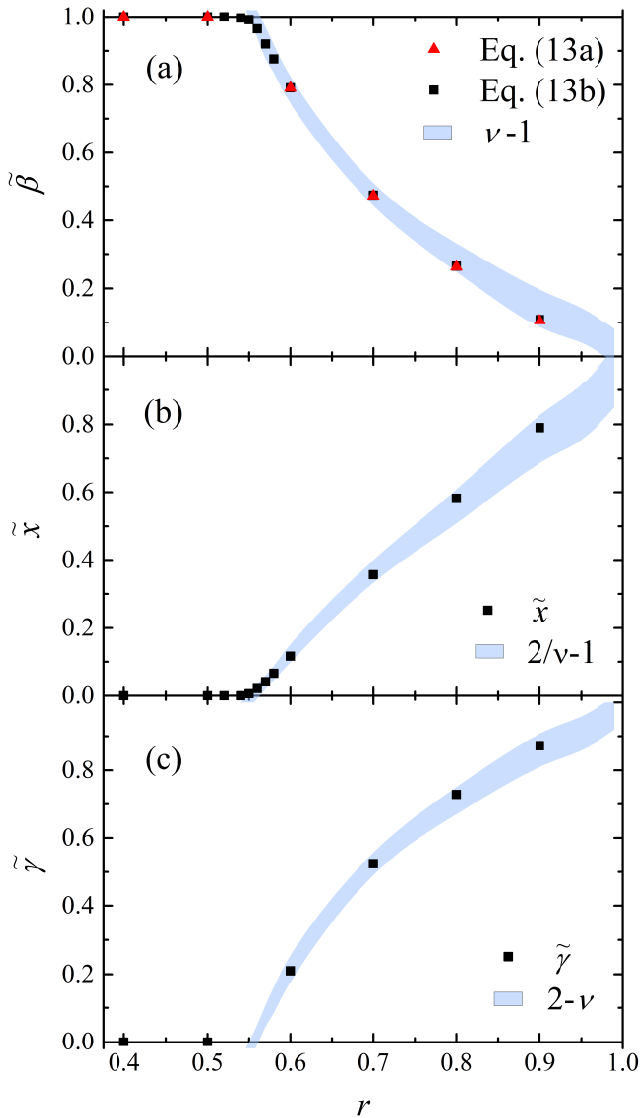}
\caption{\label{fig:exponents} (Color online)
Charge critical exponents plotted vs pseudogap exponent $r$:
(a) $\tbeta$ obtained independently from the variation of $Q_{\loc}$ with
respect to $\Ed$ and with respect to $\Gamma$, (b) $\tx$, and (c) $\tgamma$.
In all cases, the estimated nonsystematic error is smaller than the data
symbol. Shading indicates the range within which each exponent is predicted to
lie when the values of the correlation-length exponent $\nu$ from Table
\ref{tab:exponents} are inserted into Eqs.\ \eqref{charge-scaling}.}
\end{figure}

Figure \ref{fig:power-laws} illustrates the variation of $Q_{\loc}$ for the
case $r=0.6$. Irrespective of from which side the LM-ASC$_-$ boundary is
approached, $Q_{\loc}$ displays power-law variation over six decades of
$|\Gamma-\Gamma_0|$ at $\Ed=\Edz$ and over five decades of $|\Ed-\Edz|$ at
$\Gamma=\Gamma_0$.
It should be noted that these power laws reveal themselves in both phases
(unlike the power-law variation of $M_{\loc}$, which occurs only on the LM side
of the QCP). The parallel trends of the data on this log-log plot suggest that
variation of $Q_{\loc}$ with respect to $\Gamma$ and with respect to $\Ed$ is
governed by a common critical exponent $\tbeta$, i.e.,
\begin{subequations}
\label{tbeta:def}
\begin{align}
\label{tbeta:def1}
|Q_{\loc}(\Ed=\Edz)| &\propto |\dist|^{\tbeta},\\
\label{tbeta:def2}
|Q_{\loc}(\dist=0)| &\propto |\Ed-\Edz|^{\tbeta}.
\end{align}
\end{subequations}
This supposition is confirmed in Fig.\ \ref{fig:exponents}(a), which plots
values of $\tbeta$ obtained from Eqs.\ \eqref{tbeta:def1} and
\eqref{tbeta:def2} for different band exponents over the range
$0.4\le r\le 0.9$. For $r<0.4$, it proves very difficult to distinguish the
symmetric and asymmetric QCPs (which merge at $r=r^*\simeq 0.375$), while for
$r\ge 0.9$ power laws tend to become ill-defined as the system nears its upper
critical dimension \cite{Fritz:04} at $r=1$. The other striking feature of Fig.\
\ref{fig:exponents}(a) is the sharp break around $r=0.55$ between the pinned
value $\tbeta=1$ for $r\lesssim 0.55$ and the monotonic decrease of $\tbeta$
over the range $0.55\lesssim r < 1$. This decrease of $\tbeta$ points to a
variation of the impurity valence around the QCP that becomes more rapid with
increasing $r$ and presumably becomes discontinuous for $r>1$.

The $r$ dependencies of the critical exponents $\tx$ and $\tgamma$
characterizing the local charge susceptibility are plotted in Figs.\
\ref{fig:exponents}(b) and \ref{fig:exponents}(c), respectively. Each of
these exponents is positive for $r\gtrsim 0.55$, while it appears to
vanish for $r\lesssim 0.55$.

\begin{table}
\begin{tabular}{l@{\extracolsep{1em}}l@{\extracolsep{.5em}}llll}
\multicolumn{1}{c}{$r$}
  & \multicolumn{1}{c}{$\tbeta$ \eqref{tbeta:def1}}
	  & \multicolumn{1}{c}{$\tbeta$ \eqref{tbeta:def2}}
			& \multicolumn{1}{c}{$\tgamma$}
				& \multicolumn{1}{c}{$\tx$}
          & \multicolumn{1}{c}{$\nu$} \\ \hline \\[-2.1ex]
0.40 & 1.000(2)  & 1.000(1)  &          & 0.0000(1) & 4.24(4) \\
0.50 & 1.000(3)  & 1.000(2)  &          & 0.0000(2) & 2.36(4) \\
0.52 &           & 1.000(4)  &          & 0.0000(2) & 2.22(3) \\
0.54 &           & 0.997(4)  &          & 0.000(1)  & 2.08(3) \\
0.56 &           & 0.965(6)  &          & 0.021(1)  & 1.98(4) \\
0.58 &           & 0.876(5)  &          & 0.0658(6) & 1.88(4) \\
0.60 & 0.7910(6) & 0.7913(5) & 0.210(2) & 0.1164(1) & 1.77(4) \\
0.70 & 0.472(4)  & 0.474(4)  & 0.524(4) & 0.3569(4) & 1.45(3) \\
0.80 & 0.263(2)  & 0.265(2)  & 0.728(8) & 0.582(2)  & 1.29(4) \\
0.90 & 0.109(4)  & 0.105(5)  & 0.872(2) & 0.790(2)  & 1.13(6)
\end{tabular}
\caption{\label{tab:exponents}
Charge critical exponents $\tbeta$, $\tgamma$, and $\tx$, plus
correlation-length exponent $\nu$, at the particle-hole-asymmetric QCPs
of the pseudogap Anderson model for band exponents $r$ between $0.4$ and
$0.9$. Exponent $\tbeta$ was obtained independently from fits to Eqs.\
\eqref{tbeta:def1} and \eqref{tbeta:def2}. Parentheses enclose the estimated
nonsystematic error in the last digit. Each charge critical exponent
agrees to within its estimated error with the value obtained by
substituting $\nu$ into the appropriate scaling relation in Eqs.\
\eqref{charge-scaling}.}
\end{table}

Table \ref{tab:exponents} summarizes the numerical values of the three
charge critical exponents defined in Eqs.\ \eqref{charge-exponents1} and
\eqref{tbeta:def} and of $\nu$, the correlation-length exponent.
Also listed is an estimate of the non systematic error in the last decimal
place of each exponent. Exponent $\tx$ from Eq. \eqref{tx:def} generally
has the smallest error because it can be obtained from fits of $\chi_{\loc}$
over many decades of $T$. There is considerable uncertainty in the values of
$\nu$, which were obtained by interpolating from data at discrete temperatures
$T$ the value $T^*$ at which $T\chi_{\imp}(T)$ passes outside a narrow window
surrounding its critical value $X(r)$. When allowance is made for these
uncertainties, Table \ref{tab:exponents} suggests an interesting relation
among the charge critical exponents, namely,
\begin{equation}
\label{empirical-scaling}
\tbeta = 1 - \tgamma .
\end{equation}
Finally, we note that the threshold value of the band exponent $r\simeq 0.55$
seems to coincide with the point where the correlation-length exponent
passes through $\nu=2$.

Many of the empirical observations noted in the preceding paragraphs can be
understood through an extension of the scaling ansatz used
previously \cite{Ingersent:02} to explain the critical spin response. We
postulate that the singular component of the impurity free energy
takes the form given in Eq.\ \eqref{scaling-ansatz} with a generalized
definition of the nonmagnetic distance from criticality, namely, 
\begin{equation}
\label{dist:def}
\dist = (\bp - \bp_0) \cdot \nz .
\end{equation}
Here, $\nz$ is the local unit normal to the phase boundary at
$\bp_0=(U_0,\,\Edz,\,\Gamma_0)$, $h=0$ in a three-dimensional Euclidean space
of nonmagnetic couplings $\bp=(U,\,\Ed,\,\Gamma)$; the direction of
$\hat{\mathbf{n}}_0$ is chosen so that it points into the SC phase. This form
is assumed to hold for $|\bp-\bp_0|$ much smaller than the radii of curvature
of the phase boundary at $\bp_0$, in which case $|\dist|$ is just the
perpendicular distance from $\bp$ to the phase boundary.

The extended ansatz reproduces the critical spin response in Eqs.\
\eqref{spin-exponents} with exponents satisfying Eqs.\ \eqref{spin-scaling},
irrespective of whether the approach to the phase boundary is along the $U$,
$\Ed$, or $\Gamma$ axis (or along any direction in between). The ansatz also
recovers the critical charge response \cite{special_case} in Eqs.\
\eqref{charge-exponents1} and \eqref{tbeta:def}, as well as two further
power-law behaviors,
\begin{subequations}
\label{charge-exponents2}
\begin{align}
Q_{\loc}(g=0) &\propto |h|^{1/\tdelta} , \\
\chi_c(T=g=0) &\propto |h|^{-\tilde{\phi}} ,
\end{align}
\end{subequations}
with all critical exponents expressed as functions of $\nu$ and $\Delta$:
\begin{subequations}
\label{charge-scaling}
\begin{align}
\label{tbeta-scaling}
\tbeta &= \nu - 1, \\
\label{tgamma-scaling}
\tgamma &= 2 - \nu, \\
\label{tx-scaling}
\tx &= 2/\nu - 1, \\
\label{tdelta-scaling}
1/\tdelta &= (\nu - 1) / \Delta, \\
\label{tphi-scaling}
\tilde{\phi} &= (2 - \nu) / \Delta .
\end{align}
\end{subequations}
Equations \eqref{tbeta-scaling} and \eqref{tgamma-scaling} not only confirm
Eqs.\ \eqref{empirical-scaling}, but also show that since the local ``field''
$\Ed$ conjugate to the local charge enters the free energy in the same manner
as do $U$ and $\Gamma$, the charge critical exponents $\tbeta$, $\tgamma$, and
$\tx$ are functions solely of $\nu$, unlike their spin counterparts $\beta$,
$\gamma$, and $x$, which also depend on $\Delta$.

\begin{table}
\begin{tabular}{c@{\extracolsep{1.2em}}l@{\extracolsep{.8em}}llll}
\multirow{2}{*}{$r$} & \multicolumn{1}{c}{$\nu$}
  & \multicolumn{4}{c} {$\nu$ found from} \\
\cline{3-6} \\[-2ex]
   & \multicolumn{1}{c}{direct}
     & \multicolumn{1}{c}{$\tbeta$ \eqref{tbeta:def1}}
	     & \multicolumn{1}{c}{$\tbeta$ \eqref{tbeta:def2}}
				 & \multicolumn{1}{c}{$\tgamma$}
            & \multicolumn{1}{c}{$\tx$} \\ \hline \\[-2.1ex]
0.6 & 1.77(4) & 1.7910(6) & 1.7913(5) & 1.790(2) & 1.7915(6)\\
0.7 & 1.45(3) & 1.472(4)  & 1.474(2)  & 1.476(4) & 1.474(1) \\
0.8 & 1.29(4) & 1.263(2)  & 1.265(2)  & 1.272(8) & 1.264(2) \\
0.9 & 1.13(6) & 1.109(4)  & 1.105(5)  & 1.128(2) & 1.117(2)
\end{tabular}
\caption{\label{tab:scaling}
Correlation-length exponent $\nu$ at the particle-hole-asymmetric QCPs of the
pseudogap Anderson model for band exponents $r$ between $0.6$ and $0.9$, as
obtained directly and via the scaling equations \eqref{charge-scaling} from
the charge critical exponents listed in Table \ref{tab:exponents}. Except
for $r=0.9$, the various estimates of $\nu$ for a given $r$ all agree to within
the estimated nonsystematic error in the last digit of each value (enclosed in
parentheses).}
\end{table}

For all cases studied on the range $0.55 \lesssim r \le 0.9$, the directly
determined exponents $\tbeta$, $\tx$, and $\tgamma$ lie within
the bounds (represented by shaded regions in Fig.\ \ref{fig:exponents}) obtained
by inserting numerical estimates of $\nu$ into Eqs.\ \eqref{charge-scaling}.
Given the rather large uncertainties in $\nu$, a more rigorous test of the
scaling relations is provided by Table \ref{tab:scaling}, which compares
the directly determined value of $\nu$ for $0.6\le r\le 0.9$ with ones inferred
through the scaling relations from the NRG values of $\tbeta$,
$\tx$, and $\tgamma$. For each band exponent $r\le 0.8$, all values of $\nu$ agree
to within their estimated nonsystematic errors, providing strong numerical support
for the validity of Eqs.\ \eqref{charge-scaling}. We attribute the discrepancies
between the various estimates of $\nu$ for $r=0.9$ to the difficulty mentioned
above in identifying clear power-law behaviors for band exponents approaching $1$.

\begin{table}
\begin{tabular}{l@{\extracolsep{.7em}}l@{\extracolsep{.2em}}llll}
\multicolumn{1}{c}{$r$}
 
   & \multicolumn{1}{c}{$1/\tdelta$ (dir.)}
		 & \multicolumn{1}{c}{$1/\tdelta$ \eqref{tdelta-scaling}}
			 & \multicolumn{1}{c}{$\tphi$ (dir.)}
         & \multicolumn{1}{c}{$\tphi$ \eqref{tphi-scaling}}
				   & \multicolumn{1}{c}{$x$} \\
					   \hline \\[-2.1ex]
0.6 & 0.4941(5) & 0.4934(2) & 0.1306(6) & 0.1300(2) & 0.79057(6)\\
0.7 & 0.3517(3) & 0.3512(6) & 0.393(3)  & 0.390(1)  & 0.8315(1) \\
0.8 & 0.2240(7) & 0.222(2)  & 0.620(6)  & 0.619(3)  & 0.88021(7) \\
0.9 & 0.130(6)  & 0.109(3)  & 0.80(2)   & 0.820(4)  & 0.928(2)
\end{tabular}
\caption{\label{tab:h_scaling}
Exponents $\tdelta$ and $\tphi$ as determined directly (``dir.'') from
Eqs.\ \eqref{charge-exponents2} for band exponents $r$ between $0.6$
and $0.9$. Also listed are values of the same exponents inferred from
scaling equations Eqs.\ \eqref{tdelta-scaling} and \eqref{tphi-scaling},
respectively, using the best estimate of $\nu$ from Table \ref{tab:scaling}
and a value of $\Delta$ found via Eq.\ \eqref{x-scaling} from the
tabulated value of the magnetic exponent $x$. Except for $r=0.9$, the directly
determined and inferred exponents agree to within the estimated nonsystematic
error in the last digit of each value (enclosed in parentheses).}
\end{table}

Table \ref{tab:h_scaling} lists, for band exponents $0.6\le r\le 0.9$, directly
computed values of the exponents $1/\tdelta$ and $\tphi$ defined in Eqs.\
\eqref{charge-exponents2} as well the values of the same exponents predicted
from scaling Eqs.\ \eqref{tdelta-scaling} and \eqref{tphi-scaling},
respectively. For $r=0.9$, it proved difficult to obtain a robust
power-law variation of $Q_{\loc}$ with $h$, so no directly computed value is
recorded for $1/\tdelta$. The inputs to the scaling equations are (i) the value
of the correlation-length exponent $\nu$ found from $\tx$ using Eq.\
\eqref{tx-scaling} (see rightmost column of Table \ref{tab:scaling}), and
(ii) a value of the gap exponent $\Delta$ found via Eq.\ \eqref{x-scaling}
from the magnetic exponent $x$. The values of $x$ (also listed in Table
\ref{tab:h_scaling}) are either directly computed in the Anderson model
(for $r=0.7$) or obtained by refining previous results \cite{Ingersent:02}
for the pseudogap Kondo model. That the directly computed values in all cases
but one ($1/\tdelta$ for $r=0.9$) agree with their scaling predictions to within
the estimated nonsystematic errors further supports the validity of the extended
scaling ansatz contained in Eqs.\ \eqref{scaling-ansatz} and \eqref{dist:def}.

The extended scaling ansatz has implications not only for relations among
critical exponents but also for the relative magnitude of responses at
different points $\bp$ near $\bp_0$.
NRG runs performed for fixed $|\bp-\bp_0|$ but for various angles between
$\bp-\bp_0$ and the local normal $\nz$ are consistent with the hypothesis
that local spin and charge properties depend only on $\dist$ as defined in
Eq.\ \eqref{dist:def}.

For $r\lesssim 0.55$ (which is the range in which $\nu>2$), the scaling
relations in Eqs.\ \eqref{charge-scaling} predict that $\tbeta>1$ and 
$\tx, \, \tgamma<0$. In contrast, we find numerically that $\tbeta$,
$\tx$, and $\tgamma$ are pinned at trivial values of 1, 0, and 0 respectively.
In order to explain the strong deviations from scaling over this range of
band exponents, it turns out to be essential to consider the hitherto
neglected regular (analytic) parts,
\begin{equation}
F_{\imp}^{\reg} = - \frac{1}{2} \: \chi_c^{\reg} \, \dist^2
   - \frac{1}{2} \: \chi_s^{\reg} \, h^2 + \ldots ,
\end{equation}
of the total impurity free energy $F_{\imp}=F_{\imp}^{\mathrm{crit}}
+ F_{\imp}^{\reg}$. The regular terms impart a piece to $Q_{\loc}$ varying
linearly with $\dist$ (i.e., $\tbeta = 1$) and a constant local charge
susceptibility (formally corresponding to $\tx=\tgamma=0$).
For any $\nu>2$, these contributions dominate the charge responses
described by Eqs. \eqref{charge-scaling} that arise from the critical part of
the free energy. The condition $\nu>2$ does not preclude a divergent local
spin susceptibility, which depends not only on $\nu$ but also the gap
exponent $\Delta$. Indeed, nontrivial critical behavior in the spin sector
persists for $r\to 0^+$, in which limit there is a divergent correlation-length
exponent $\nu\simeq 1/r$ (Refs.\ \onlinecite{Fritz:04} and
\onlinecite{Kircan:04}).

\begin{figure}
\includegraphics[width=\columnwidth]{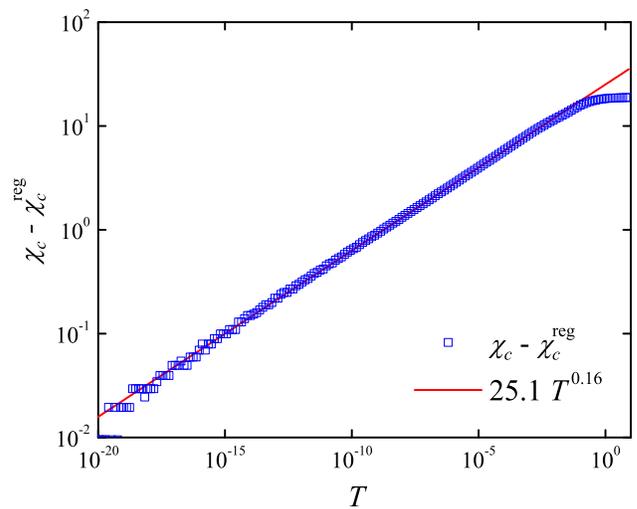}
\caption{\label{fig:chi_c-correction} (Color online)
Temperature-dependent part $\chi_c - \chi_c^{\reg}$ of the local charge
susceptibility for band exponent $r=0.5$. The line fitted through the NRG data
points corresponds to $\chi_c - \chi_c^{\reg}\propto T^{0.16}$.}
\end{figure}

It should be pointed out that a nonanalytic charge response, albeit subleading,
is still present in the range of band exponents where $\nu>2$. This is
illustrated in Fig.\ \ref{fig:chi_c-correction}, a log-log plot of
$\chi_c(T)-\chi_c^{\reg}$ [where $\chi_c^{\reg}\equiv\chi_c(T\to 0)]$ versus
temperature for the representative case $r=0.5$. An empirical fit
$\chi_c-\chi_c^{\reg}\propto T^{0.16}$ is in close agreement with the
expectation based on Eq.\ \eqref{tx-scaling} of a temperature exponent
$1-2/\nu = 0.15(3)$. 

\section{Discussion}
\label{sec:discuss}

This work has shed light on the critical local charge response found previously
near the Kondo-destruction quantum critical point (QCP) in the pseudogap
Anderson impurity model away from particle-hole symmetry \cite{Pixley:12}.
The ``field'' conjugate to the local charge (i.e., the impurity occupancy) is
the impurity level energy $\Ed$. Changing $\Ed$ does not destroy or restore the
SU(2) spin-rotation invariance that distinguishes the model's strong-coupling
phase from its broken-symmetry local-moment phase. For this reason, $\Ed$ joins
other model couplings, such as the interaction strength $U$ and the
hybridization width $\Gamma$, whose collective deviation $\dist$ from the phase
boundary enters an ansatz [Eq.\ \eqref{scaling-ansatz}] for the free energy in
the scaling combination $\dist/T^{1/\nu}$, distinct from the
$|h|/T^{\Delta/\nu}$ scaling of the local magnetic field \cite{Kondo_ansatz}.
First and second partial derivatives of the free-energy with respect to $\Ed$
exhibit power-law variations with exponents $\tbeta$, $\tgamma$, $\tdelta$,
$\tphi$, and $\tx$ that depend on $\nu$, but (apart from $\tdelta$ and $\tphi$)
are independent of $\Delta$. Presumably, the corresponding partial derivatives of
the free energy with respect to $U$ and $\Gamma$ would be described by the same
set of exponents.

In all cases studied numerically in this work, the local charge response at the
QCP has proved to be less singular than the local spin response. However, it is
straightforward to come up with an example where the reverse ordering holds.
Interchange of spin and charge degrees of freedom maps the $U>0$ Anderson model
in zero magnetic field to a $U<0$ Anderson model at particle-hole symmetry. In
the presence of a pseudogapped density of states described by exponent $0<r<1$,
this negative-$U$ Anderson model must have a QCP between strong-coupling and
local-charge phases \cite{Cheng:13} at which the local charge response is governed
by critical exponents $\beta$, $\gamma$, $\delta$, and $x$, while the local spin
response is weaker and described by critical exponents $\tbeta$, $\tgamma$,
$\tdelta$, $\tphi$, and $\tx$.

What does seem intuitively reasonable is that the response to the
order-parameter field is more singular than that to other perturbations of the
system. Indeed, one can argue that this should be true at any interacting
QCP described by the scaling ansatz Eq.\ \eqref{scaling-ansatz}, examples of
which have been identified in a number of other quantum impurity
models \cite{Vojta:05,Glossop:05,Vojta:06,Chung:07,Cheng:09,Pixley:13}. At such
a QCP, the response to the order-parameter field will be the most singular
response provided that the gap exponent satisfies $\Delta>1$, a condition that
can be shown using Eqs.\ \eqref{spin-scaling} and \eqref{x-scaling} [all
derived from Eq.\ \eqref{scaling-ansatz}] to be equivalent to $\beta+\gamma>1$.
Since any interacting QCP is expected to satisfy $\beta>0$ (describing a
continuous power-law rise of the order parameter) and $\gamma\ge 1$ ($\gamma=1$
being the mean-field value), $\beta+\gamma>1$ should be satisfied quite
generally.

In summary, we have provided a unified picture of critical spin and charge
responses at quantum critical points in the particle-hole-asymmetric pseudogap
Anderson Hamiltonian, a toy model for investigating critical Kondo destruction
at mixed valence. All critical exponents have been related to just two
underlying exponents: the correlation-length exponent $\nu$ and the gap exponent
$\Delta$. The charge susceptibility diverges at the transition provided $\nu<2$,
while for $\nu>2$ the local charge response is regular with nonanalytic
corrections. We have argued that nonanalytic responses to non-symmetry-breaking
fields are a generic feature of interacting QCPs in quantum impurity models,
although such responses should be less singular than those to a field breaking
the symmetry that distinguishes the phases on either side of the QCP.

\section{Acknowledgments}

We acknowledge useful conversations with J. Pixley and Q. Si. This work
has been supported by NSF Grant No.\ DMR-1107814.


\begin{thebibliography}{30}
\bibitem{nomenclature}
In the literature on impurity quantum phase transitions, $\Delta/\nu$ is
usually written as $(1+x)/2$ or $1-\eta/2$, where $\eta=1-x\ge 0$ is the
anomalous magnetic exponent.
\bibitem{Kondo_ansatz}
In the limit of unit impurity occupancy, the pseudogap Anderson model can be
mapped at low energies to a pseudogap Kondo model with an exchange coupling
$J$ and a potential scattering $V$. In this Kondo model, we expect that
the critical part of the impurity free energy satisfies the scaling ansatz in
Eq.\ \eqref{scaling-ansatz} with $|g|$ being the length of the perpendicular to
the phase boundary [parametrized as $J_0 = J_c(V_0)$] that passes through the
point $(J,V)$ representing the bare model parameters.
\end{thebibliography}
\end{document}